\documentclass[11pt]{article}

\usepackage{amssymb,amsmath}

\usepackage[top=2.0cm,bottom=2.0cm,left=3.3 cm,right=3.3cm,includehead,includefoot]{geometry}

\newenvironment{prf}{\noindent {\it Proof} \ }{\hfill $\Box$}

\newcommand{\p}{\partial}
\newcommand{\D}{\mathcal{D}}
\newcommand{\nn}{\notag}
\newcommand{\lm}{\lambda}
\newcommand{\res}{\textrm{res}}

\usepackage{amsthm}
\newtheorem*{thm}{Main Theorem}
\newtheorem*{lemma}{Lemma}
\newtheorem*{conj}{Conjecture}

\begin{document}

\title{Central Invariants of the Constrained KP Hierarchies}
\author{Si-Qi Liu, Youjin Zhang, Xu Zhou \\
\small Department of Mathematical Sciences, Tsinghua University, \\
\small Beijing, 100084, China}
\maketitle

\begin{abstract}
We compute the central invariants of the bihamiltonian structures of the constrained KP hierarchies, 
and show that these integrable hierarchies are topological deformations of their hydrodynamic limits.
\end{abstract}

\parskip 1ex

\section{Introduction}

The notion of central invariants of a semisimple bihamiltonian structure which possesses hydrodynamic limit was introduced in \cite{DLZ1, LZ1}. It was proved in \cite{DLZ1} that such invariants completely characterize the equivalence classes of the infinitesimal deformations of a semisimple bihamiltonian structures of hydrodynamic type under Miura type transformations. It was also conjectured in  \cite{LZ1, DLZ1} that for any given semisimple bihamiltonian structure of hydrodynamic type and a set of central invariants there exists a deformation of the bihamiltonian structure. In order to prove the conjecture, one needs to verify the triviality of the associated third bihamiltonian cohomolgy that is introduced in \cite{DZ}. In \cite{LZ2} the first two authors of the present paper developed an approach to compute the bihamiltonian cohomologies of a semisimple 
bihamiltonian structure, and proved the conjecture for the one component case.
In a recent preprint \cite{CPS2}, by using the approach of \cite{LZ2} and by further developing the computing technique, Carlet, Posthuma and Shadrin proved the conjecture for the general case. So for a given bihamiltonian structure, the knowledge of its central invariants is crucial in order to understand its properties and to study the relationship of the 
associated integrable hierarchy of evolutionary PDEs with different research fields in mathematical physics. For example, for a bihamiltonian integrable hierarchy that controls a cohomological field theory which is associated to a semisimple Frobenius manifold, we know from \cite{LZ1, DLZ1, Zh} that the central invariants of its bihamiltonian structure must be equal to $\frac1{24}$. Such a bihamiltonian integrable hierarchy is called the topological deformation of its hydrodynamic limit \cite{DZ}.  This property of central invariants provides us a useful criterion to find candidates of bihamiltonian integrable hierarchies that are related to cohomological field theories.   

Since the definition of the central invariants of a semisimple bihamiltonian structure involves representation of 
the bihamiltonian structure in terms of its canonical coordinates, it is actually quite nontrivial to compute these invariants. In \cite{DLZ2}, the central invariants of the bihamiltonian structures associate to the Drinfeld-Sokolov hierarchies \cite{DS} are obtained. These bihamiltonian integrable hierarchies are labeled by the untwisted affine Lie algebras.
For the integrable hierarchies that are associated to the untwisted affine Lie algebras of ADE type, the central invariants of their bihamiltonian structures are equal to $\frac1{24}$. It is shown in \cite{FJR} that certain particular tau functions of these integrable hierarchies yield the total descendant potentials of the FJRW invariants of the singularities of ADE type.  It is further shown in \cite{LRZ} that these Drinfeld-Sokolov hierarchies admit
certain finite symmetry group actions, and the invariant flows w.r.t. these group actions coincide with the Drinfeld-Sokolov hierarchies associated to the untwisted affine Lie algebras of BCFG type. It is further proved in \cite{LRZ} that these BCFG Drinfeld-Sokolov hierarchies control the invariant sectors of the FJRW 
cohomological field theory for singularities of  ADE type under certain symmetry group actions. A natural question then arises: whether there are similar relations between the Drinfeld-Sokolov hierarchies associated to 
the twisted affine Lie algebras and cohomological field theories. Let us note that, instead of a bihamiltonian structure,  a Drinfeld-Sokolov hierarchy which is associated to a twisted affine Lie algebra only possesses one Hamiltonian structure \cite{DS, DLZ2}. So the first step of our approach to reveal its relationship with a certain cohomological field theory is to represent it as the reduction of a bihamiltonian integrable hierarchy, and we
require that the central invariants of the bihamiltonian structure are equal to $\frac1{24}$.

For the Drinfeld-Sokolov hierarchy associated to the twisted affine Lie algebras of type $A^{(2)}_{2n-1}$,
we found that it can be obtained as the invariant flows of a certain generalized Drinfeld-Sokolov hierarchy
associated to the untwisted affine Lie algebra of type $A^{(1)}_{2n-1}$. Such an integrable hierarchy was 
systematically studied in the beginning of the 90's of the last century, and bears the name of a constrained KP 
hierarchy \cite{BX, CL, Ch1, KSS, Kr}. It possesses a bihamiltonian structure which admits hydrodynamic limit \cite{Ch2, OS}. The purpose 
of the present paper is to compute the central invariants of this bihamiltonian structure, and we obtain the following 
theorem.

\begin{thm} \label{thm-1}
The central invariants of the bihamiltonian structures of the constrained KP hierarchies are equal 
to $\frac{1}{24}$.	 
\end{thm}

The paper is organized as follows. In Sec.\,\ref{sec-2} we give the definition of the constrained KP hierarchies, and present their bihamiltonian structures. In Sec.\,\ref{sec-3} we recall the definition of central invariants of a semisimple bihamiltonian structure. 
In Sec.\,\ref{sec-4} we compute explicitly the central invariants of the constrained KP hierarchies, and prove the Main Theorem. In the last section we present a conjecture 
that relates the constrained KP hierarchies with the topological deformations of the principal hierarchies associated to a certain class of Frobenius manifolds.
 
\section{The constrained KP hierarchies and their bihamiltonian structures }\label{sec-2}

Let us denote by $\D$  the ring of pseudo-differential operators of the form $\sum_{k\le m} f_k \p^k$,
where $f_k: x\mapsto f_k(x)$ are smooth functions on $\mathbb{R}$ and $m$ is an integer.
For any two pseudo-differential operators 
\[A=\sum_{k\le m_1} f_k \p^k,\quad B=\sum_{k\le m_2} g_k \p^k\]
of $\mathcal{D}$, their product is defined as follows:
\[ A B=\sum_{k\le m_1}\sum_{j\le m_2}\sum_{l\ge 0}\binom{k}{l} f_k \frac{\p^l g_j}{\p x^l} \p^{k+j-l},
\quad \binom{k}{l}=\frac{k(k-1)\dots (k-l+1)}{l!}.\]

For a fixed non-negative integer $n$, we consider the pseudo-differential operator $L$ of the following form:
\begin{equation}\label{zh-2}
L = \partial ^{n+1} + v_n\partial ^{n-1} + \cdots + v_2\partial + v_1 + (\p-w)^{-1} u,
\end{equation}
where 
\[ (\p-w)^{-1}=a_1 \p^{-1}+a_2 \p^{-2}+\dots\]
is uniquely determined by the identity 
\[ (\p-w)(a_1 \p^{-1}+a_2 \p^{-2}+\dots)=1.\]
We denote the differential operator part of a pseudo-differential operator $P$ by $P_+$. Then the $(n+1)$-constrained KP hierarchy can be represented by the following Lax equations:
\begin{equation}\label{zh-1}
\frac{\partial L}{\partial t_k} = [(L^{k/(n+1)})_+ , L],\quad k\ge 1.
\end{equation}
For example, the first flow of the constrained KP hierarchy \eqref{zh-1} is the translation along the spatial variable $x$. The second flow of the hierarchy for the $1$-constrained KP hierarchy is given by
\begin{align}
&\frac{\p u}{\p t_2}=2 w u_x+2 u w_x-u_{xx},\nn\\
& \frac{\p w}{\p t_2}=2 u_x+2 w w_x+w_{xx},\nn
\end{align}
and that of the 2-constrained KP hierarchy has the expression
\begin{align}
&\frac{\p v_1}{\p t_2}=2 u_x,\quad 
\frac{\p u}{\p t_2}=2 w u_x+2 u w_x-u_{xx},\nn\\
&\frac{\p w}{\p t_2}=2 w w_x+v_{1,x}+w_{xx}.\nn
\end{align}
Note that in the literature the constrained KP hierarchy is usually represented 
by the equations
\begin{align*}
&\frac{\p \mathcal{L}}{\p t_k}=[(\mathcal{L}^{\frac{k}{n+1}})_+, \mathcal{L}],\quad k\ge 1,\\
&\frac{\p q}{\p t_k}=(\mathcal{L}^{\frac{k}{n+1}})_+ q,\quad 
\frac{\p r}{\p t_k}=-(\mathcal{L}^{\frac{k}{n+1}})_+^* r, \quad k\ge 1.
\end{align*}
Here the pseudo-differential operator $\mathcal{L}$ has the form
\[ \mathcal{L} = \partial ^{n+1} + v_n\partial ^{n-1} + \cdots + v_2\partial + v_1 + q \p^{-1} r.
\]
These flows are related to the ones  defined by Lax equations \eqref{zh-1} by the following 
transformations of the dependent variables
\[
u=q r, \qquad w = \frac{q_x}{q} .
\]

As it was shown in \cite{Ch2, OS}, the constrained KP hierarchy \eqref{zh-1} possesses a bihamiltonian structure. In order to present this bihamiltonian structure, let us first introduce
some notations. Denote 
\[ B=L_+=\p^{n+1}+v_n \p^{n-1}+\dots+v_1.\]
We define the variational derivative of a functional 
\[F=\int f(\mathbf{v}, \mathbf{v}_x, \dots) dx,\quad \mathbf{v}=(v_1, v_2,\dots, v_n, w,u)\]
with respect to the pseudo-differential operator $L$ by
\begin{align}\label{zh-3}
\frac{\delta F}{\delta L} := \frac{\delta F}{\delta B} + \frac{\delta F}{\delta u} +\frac{\delta F}{\delta w} \frac{1}{u} (\partial - w), 
\end{align}
where the variational derivative of $F$ with respect to the differential operator $B$ is defined as usual \cite{Di} by
\[
\frac{\delta F}{\delta B} = \sum_{i=1}^n \partial^{-i} \frac{\delta F}{\delta v_i}.
\]
Then it is easy to verify that the variation of the functional $F$, defined by
\[
\delta F = \int \left( \sum_{i=1}^n \frac{\delta F}{\delta v_i(x)}\delta v_i(x)  + \frac{\delta F}{\delta u(x)}\delta u(x) +\frac{\delta F}{\delta w(x)}\delta w(x) \right) dx, 
\]
can be represented as
\[
\int \mathrm{res} \left( \frac{\delta F}{\delta L} \delta L \right) dx, 
\]
where the residue of a pseudo-differential operator is defined by
\[
\mathrm{res} ( \sum_{i \leq m} a_i \partial ^{i} ) = a_{-1} .
\]

For two functionals 
\[F=\int f(\mathbf{v}, \mathbf{v}_x, \dots) dx,\quad G=\int g(\mathbf{v}, \mathbf{v}_x, \dots) dx,\]
denote their variational derivatives w.r.t. $L$ by 
\[
X=\frac{\delta F}{\delta L},\quad Y=\frac{\delta G}{\delta L}.
\]
Then by a straightforward computation, we can represent the two compatible Poisson brackets given in \cite{Ch2, OS} for the constrained KP hierarchy as follows:
\begin{align}
&\{F,G\}_1 = \int \mathrm{res} \left([L,X_+] Y-[L,X]_+Y\right) \; dx ,\label{ds-bh-1}\\
&\{F,G\}_2 = \int \mathrm{res} \left( (LY)_+ LX - (YL)_+ XL + \frac{1}{n+1} X[L,K_Y] \right) \; dx.\label{ds-bh-2} 
\end{align}
Here $K_Y$ is given by the differential polynomial $\partial ^{-1}_x \mathrm{res}([L,Y])$. We note that the residue of the commutator of two pseudo-differential operators is always a total derivative with respect to $x\,$, hence $K_Y$ is well-defined. 

Define the Hamiltonians
\begin{equation}\label{ds-bh-3}
H_k=\int h_k(\mathbf{v}, \mathbf{v}_x, \dots) dx,\quad k\ge -n
\end{equation}
with the densities 
\[ h_k=\frac{n+1}{k+n+1} \mathrm{res}L^{\frac{k+n+1}{n+1}}.\]
Then it follows from \cite{Ch2, OS} that the constrained KP hierarchy \eqref{zh-1} has the following bihamiltonian 
representation:
\begin{equation}\label{ds-bh-5}
\frac{\p \mathbf{v}}{\p t_k}=\{\mathbf{v}(x), H_k\}_1=\{\mathbf{v}(x), H_{k-n-1}\}_2,\quad  k\ge 1.
\end{equation}

We remark here that the original expressions of the two Poisson brackets that are given in \cite{Ch2, OS} are quite long. 
However, with our introduction of the notion \eqref{zh-3} of  the variational derivative of the functional 
w.r.t. the pseudo-differential operator $L$, we are able to obtain the above simple expressions 
of the two Poisson brackets. This representation of the bihamiltonian structure
is essential for our computation of its central invariants that is given in Sec.\,\ref{sec-4}.

\section{Definition of the central invariants}\label{sec-3}

We consider bihamiltonian structures of the form
\begin{equation}\label{biham}
\{F, G\}_a=\int \frac{\delta F}{\delta w_i} \mathcal{P}_a^{ij} \frac{\delta G}{\delta w_j} dx,\quad a=1,2, 
\end{equation}
where the local functionals $F, G$ are defined on the jet space of a $n$-dimensional manifold 
$M$ with local coordinates $w_1,\dots, w_n$,  the Hamiltonian operators $P_1^{ij}, P_2^{ij}$
have the form
\[ \mathcal{P}_a^{ij}=g^{ij}_a(w) \p_x+\Gamma_{k;a}^{ij}(w) w_{k,x}+\sum_{k\ge 1} \epsilon^k
 A_{k,l;a}^{ij}(w;w_x,\cdots w^{(l)})\p_x^{(k-l+1)}
\]
for $a=1, 2$. Here the matrices $(g^{ij}_a),  a=1,2$ are non-degenerate symmetric, their elements are smooth functions of $w_1,\dots, w_n$. The functions $A^{ij}_{k,l;a}$
are homogeneous polynomials of $w_{k,x},\dots, \p_x^l w_k$ of degree $l$ with coefficients depending smoothly on $w_1,\dots, w_n$. Here we define the degrees of the jet variables by 
\[ \deg \p_x^k w_j=k, \quad k\ge 0,\ j=1,\dots, n.\]   
The semisimplicity of the bihamiltonian structure requires that the root $\lm_1(w), 
\dots, \lm_n(w)$ of the characteristic polynomial
\[\det(g^{ij}_2-\lm g^{ij}_1)\]
form a system of local coordinates of the manifold $M$ near a generic point of $M$ which are called the canonical coordinates
of the bihamiltonian structure. In these coordinates, the matrices $(g^{ij}_a(\lm))$ have diagonal forms
\[ g^{ij}_1(\lm)=f^i(\lm)\delta_{ij},\quad 
g^{ij}_2(\lm)=\lm_i f^i(\lm)\delta_{ij},\]
where $\delta_{ij}$ is the Kronecker delta function, and 
\[g^{ij}_a(\lm)=\frac{\p \lm_i}{\p w_k} g_a^{kl}(w)\frac{\p \lm_j}{\p w_l},\quad a=1, 2.\]

Denote 
\[P_a^{ij}(\lambda) =\frac{\p \lm_i}{\p w_k} A^{kl}_{1,0;a}(w)\frac{\p \lm_j}{\p w_l},\quad
Q_a^{ij}(\lambda) =\frac{\p \lm_i}{\p w_k} A^{kl}_{2,0;a}(w)\frac{\p \lm_j}{\p w_l},\]
then the central invariants $c_1(\lm), \dots, c_n(\lm)$ of the bihamiltonian structure \eqref{biham} are defined by \cite{DLZ1}
\begin{equation} \label{CenInv}
c_i(\lambda) = \frac{1}{3(f^i(\lambda))^2}\left[ Q^{ii}(\lambda)- \lambda^i Q^{ii}_1(\lambda) + \sum_{k \neq i} \frac{(P_2^{ki}(\lambda)-\lambda^i P_1^{ki}(\lambda))^2}{f^k(\lambda)(\lambda^k - \lambda^i)} \right].
\end{equation}

Now let us introduce the dispersion parameter $\epsilon$ to the constrained KP hierarchy \eqref{zh-1} and its bihamiltonian 
structure \eqref{ds-bh-1}--\eqref{ds-bh-5} by the following rescaling 
\[ \frac{\p}{\p t_k}\to \epsilon \frac{\p}{\p t_k},\quad \frac{\p}{\p x}\to \epsilon \frac{\p}{\p x},\quad \p\to D=\epsilon \p.\]
Then for two pseudo-differential operators 
\[A=\sum_{k\le m_1} f_k D^k,\quad B=\sum_{k\le m_2} g_k D^k,\]
their product is modified as
\[ A B=\sum_{k\le m_1}\sum_{j\le m_2}\sum_{l\ge 0}\epsilon^l \binom{k}{l} f_k \frac{\p^l g_j}{\p x^l} \p^{k+j-l}.\]
 After the rescaling, the constrained KP hierarchy \eqref{zh-1} takes the form
\begin{equation}\label{jw-1}
\frac{\partial L}{\partial t_k} = \frac1{\epsilon} [(L^{k/(n+1)})_+ , L],\quad k\ge 1,
\end{equation}
where the Lax operator is given by 
\begin{equation}\label{ConKP}
L = D^{n+1} + v_n D^{n-1} + \cdots + v_2 D + v_1 + (D-w)^{-1}u.
\end{equation}
The associated compatible pair of Poisson brackets \eqref{ds-bh-1}, \eqref{ds-bh-2} 
is modified as 
\begin{align}
&\{F,G\}_1 = \frac{1}{\epsilon} \int \mathrm{res} \left([L,X_+] Y-[L,X]_+Y\right) \; dx ,\label{PoBkt1}\\
&\{F,G\}_2 = \frac{1}{\epsilon} \int \mathrm{res} \left( (LY)_+ LX - (YL)_+ XL + \frac{1}{n+1} X[L,K_Y] \right) dx ,\label{PoBkt2}
\end{align}
with $ K_Y = \epsilon^{-1} \p_x^{-1} \mathrm{res}([L,Y])$. Here and in what follows the residue of a pseudo-differential operator is defined to be the coefficient of $D^{-1}$.
It is easy to see that the bihamiltonian structure \eqref{PoBkt1}, \eqref{PoBkt2} for the constrained KP hierarchy \eqref{zh-1} has the form \eqref{biham}, so we can define its central invariants by using the formulae given in \eqref{CenInv}.
In the next section we are going to show that all these invariants are 
equal to $\frac1{24}$, which proves the Main Theorem.

\section{Proof of the Main Theorem}\label{sec-4}

In order to compute the central invariants for the bihamiltonian structure of the constrained KP hierarchy, we first recall the notion of the symbol of a pseudo-differential operator and 
the star product between two such symbols, see \cite{DLZ2} for more details. 

For a pseudo-differential operator $A=\sum_{i\leq n} a_i(x)D^i$, its symbol $\hat A(x, p)$ is defined as
\[
\hat A(x,p) = \sum_{i\leq n} a_i(x)p^i. 
\]
We will also denote it as $\hat A(p)$ or simply as $\hat A$ when its dependence on the spatial variable $x$ and on the parameter $p$ is clear from the context.  
The positive part of the symbol $A$ and the residue of it are defined by
\[
\hat{A}(p)_+ =\sum_{0\le i\le n} a_i(x) p^i=\frac{1}{2 \pi i} \oint \frac{\hat A (q)}{q - p}dq,
\quad \res \hat A=a_{-1}=\frac{1}{2 \pi i} \oint \hat A (q) dq.
\]
Here the integration is taken along a circle $\gamma=\{q\in \mathbb{C}\,|\, |q|=r\}$ with anticlockwise orientation and $r>|p|$.
The star product of two symbols $\hat A_1$, $\hat A_2$ is defined by
\begin{align*}
\hat A_1(x,p) \star \hat A_2(x,p) :=& e^{\epsilon \frac{\partial^2}{\partial p \partial x'}} A_1(x, p)A_2(x', p')|_{x'=x,\, p'=p} \\
=& \sum_{k=0}^\infty \frac{\epsilon^k}{k!}\partial_p^k\hat A_1(x,p) \partial_x^k\hat A_2(x,p). 
\end{align*}
It is easy to see that, giving two pseudo-differential operators $A_1$ and $A_2$, the symbol of their product $A_1 A_2$ is just the star  product of the corresponding symbols, i.e.
\[
\widehat{A_1 A_2} = \hat A_1 \star \hat A_2 .
\]
It follows from the above definition that the symbol of the Lax operator $L$ has the expression
\begin{align*}
\hat{L}(p)
&= p^{n+1} + v_n(x) p^{n-1} + \cdots + v_2(x) p + v_1(x) + \left((D-w(x))^{-1}\right)\hat{\ }\star u,
\end{align*}
where the symbol 
\[ \left((D-w)^{-1}\right)\hat{\ }=\sum_{k\ge 1} \frac{b_k}{(p-w)^k}\]
is determined by the following recursive relations:
\[b_1=1, \quad b_2=0,  \quad b_{k+1}=-\epsilon (k-1) b_{k-1} w_x-\epsilon b_{k,x},\quad k\ge 2.\]
Thus the symbol of $L$ can be represented as 
\begin{equation}\label{SymL}
\hat{L}(p) = \lambda(x, p) +  \sum_{k=1}^{\infty} \epsilon^k \lambda_k(x, p)
\end{equation}
with the leading term
\begin{equation}
\lambda(x, p) =  p^{n+1} + v_n(x) p^{n-1} + \cdots + v_2(x) p + v_1(x) + \frac{u(x)}{p-w(x)}.
\end{equation}

Now we proceed to compute the coefficients $P_a^{ij}$, $Q_a^{ij}$ and $f^i$ of the bihamiltonian structure \eqref{PoBkt1}, \eqref{PoBkt2}. For a fix $y\in\mathbb{R}$, 
we can regard $v_1(y), \dots, v_n(y), w(y), u(y)$ as local functionals. We introduce 
a generating function $F(y)$ by
\begin{equation}
F(y)=\lambda (y, \xi) - \xi^{n+1}= \int \left( \sum_{k=1}^n v_k(x) \xi^{k-1} + \frac{u(x)}{\xi-w(x)}\right) \delta(x-y)\,dx.
\end{equation}
Here $\delta(x-y)$ is the Dirac delta function.
In a similar way we introduce functional  $G(z)=\lambda(z,\zeta) - \zeta^{n+1}$. We are to compute the Poisson brackets of these two functionals $\{F(y),G(z)\}_a (a=1,2)$
which we will also denote by $\{\lambda (y,\xi) ,\lambda(z,\zeta)\}_a$.

By using our definition \eqref{zh-3} of the variational derivative w.r.t. the Lax operator $L$ we have
\begin{align*}
X :&= \frac{\delta F(y)}{\delta L} \\
&= \sum_{k=1}^n D^{-k} \xi^{k-1} \delta(x-y) + \frac{\delta(x-y)}{\xi-w(x)} + \frac{ \delta(x-y)}{(\xi-w(x))^2}(D-w(x)) \\
&= \sum_{k=-1}^n D^{-k}a_k(x) - \epsilon a_{-1}'(x),
\end{align*}
where 
\begin{align}
a_{-1}(x)&=\frac{\delta(x-y)}{(\xi-w(x))^2},\nn\\
a_0(x)&=\frac{\delta(x-y)}{\xi-w(x)}-\frac{\delta(x-y)w(x)}{(\xi-w(x))^2} ,\nn\\
a_k(x)&=\xi^{k-1}\delta(x-y),\quad k=1,2,\dots, n.\nn
\end{align} 
The variational derivative of the functional $G(z)$ w.r.t. the Lax operator $L$ 
has a similar expression
\begin{align*}
Y 
&:= \frac{\delta G(z)}{\delta L} = \sum_{k=-1}^n D^{-k}b_k(x)- \epsilon b_{-1}'(x),\ \textrm{with}\ b_k(x)=a_k(x)|_{\xi\to \zeta,\, y\to z}.
\end{align*}
Let us denote 
\[f(p) = \sum_{i=-1}^n \frac{a_i(x)}{p^i},\quad g(p) = \sum_{i=-1}^n \frac{b_i(x)}{p^i}\]
and $a(x) = a_{-1}(x), b(x) = b_{-1}(x)$.
Then it is easy to see that the symbols of $X$ and $Y$ have the expressions
\begin{align}
&\hat{X} = \sum_{k=-1}^n p^{-k} \star a_k(x) - \epsilon a'(x) = \sum_{k \geq 0} \frac{\epsilon^k}{k!}\partial_p^k\partial_x^k f(p) - \epsilon a'(x), \label{SymX}\\
&\hat{Y} = \sum_{k=-1}^n p^{-k} \star b_k(x) - \epsilon b'(x) = \sum_{k \geq 0} \frac{\epsilon^k}{k!}\partial_p^k\partial_x^k g(p) - \epsilon b'(x).\label{SymY}
\end{align}
Now we are ready to compute the Poisson brackets \eqref{PoBkt1}, \eqref{PoBkt2}. We have
\begin{align*}
&\{\lambda (y,\xi) ,\lambda(z,\zeta)\}_1 = \frac{1}{\epsilon} \int \mathrm{res} \left([L,X_+] Y-[L,X]_+Y\right) dx ,
 \\
=& \frac{1}{\epsilon} \int \mathrm{res} \left( [L,Y]_+ X + Y_+ [L,X] \right) dx \\
=& \frac{1}{\epsilon} \int \mathrm{res} \left( \left( \hat{L}(p) \star \hat{Y}(p) - \hat{Y}(p) \star \hat{L}(p) \right) _+ \star \hat{X}(p) \right) dx \\
 &\quad +\frac{1}{\epsilon} \int \mathrm{res} \left( \hat{Y}(p)_+ \star \left( \hat{L}(p) \star \hat{X}(p) - \hat{X}(p) \star \hat{L}(p) \right) \right) dx \\
=& \frac{1}{\epsilon} \int dx \oint \frac{dp}{2 \pi i} \left( \oint \frac{\hat{L}(q) \star \hat{Y}(q) - \hat{Y}(q) \star \hat{L}(q)}{q-p} \frac{dq}{2 \pi i} \right) \star \hat{X}(p) \\
 &\quad +\frac{1}{\epsilon} \int dx \oint \frac{dp}{2 \pi i} \left( \oint \frac{\hat{Y}(q)}{q-p} \frac{dq}{2 \pi i}  \right) \star \left( \hat{L}(p) \star \hat{X}(p) - \hat{X}(p) \star \hat{L}(p) \right).
\end{align*}
Since the definition \eqref{CenInv} of the central invariants does not involve the terms  in the expression of the bihamiltonian structure that contain derivatives of the variables $u$, $w$ and $v_i$'s with respect to $x$, we see that the second term in the r.h.s. of \eqref{SymL} does not 
contribute to the central invariants. Thus in the above expression of the Poisson bracket, we can replace $\hat{L}(p)$ by $\lambda(x,p)$ which we also denote as $\lambda(p)$. Similarly, for the second Poisson bracket we have 
\begin{align*}
&\{\lambda (y,\xi) ,\lambda(z,\zeta)\}_2  \\
=& \frac{1}{\epsilon} \int \mathrm{res} \left( (LY)_+ LX - (YL)_+ XL + \frac{1}{n+1} X[L,K_Y] \right) dx \\
=& \frac{1}{\epsilon} \int \mathrm{res} \left( (\hat{L}(p) \star \hat{Y}(p))_+ \star \hat{L}(p) \star \hat{X}(p) \right) dx \\
 &\quad - \frac{1}{\epsilon} \int \mathrm{res} \left( (\hat{Y}(p) \star \hat{L}(p))_+ \star \hat{X}(p) \star \hat{L}(p) \right) dx \\
 &\quad + \frac{1}{\epsilon (n+1)} \int \mathrm{res} \left( \hat{X}(p) \star (\hat{L}(p) \star K_Y - K_Y \star \hat{L}(p)) \right) dx \\
= &\frac{1}{\epsilon} \int dx \oint \frac{dp}{2 \pi i} \left( \oint \frac{\hat{L}(q) \star \hat{Y}(q)}{q-p} \frac{dq}{2 \pi i} \right) \star \hat{L}(p) \star \hat{X}(p) \\
 &\quad - \frac{1}{\epsilon} \int dx \oint \frac{dp}{2 \pi i} \left( \oint \frac{\hat{Y}(q) \star \hat{L}(q)}{q-p} \frac{dq}{2 \pi i} \right) \star \hat{X}(p) \star \hat{L}(p) \\
 &\quad + \frac{1}{\epsilon (n+1)} \int dx \oint \frac{dp}{2 \pi i} \left( \hat{X}(p) \star (\hat{L}(p) \star K_Y - K_Y \star \hat{L}(p)) \right).
\end{align*}
In the above formula, the function $K_Y$ can be represented as  
\begin{align}
K_Y & = \epsilon^{-1}\p_x^{-1} \mathrm{res}([L,Y]) \nn\\
    & = \epsilon^{-1}\p_x^{-1} \mathrm{res} \left( \hat{L}(q) \star \hat{Y}(q) - \hat{Y}(q) \star \hat{L}(q) \right) \nn\\
    & = \oint \frac{dq}{2 \pi i} \sum_{k=1}^{\infty} \frac{\epsilon^{k-1}}{k!} \p_{p}^{k}\hat{L}(q) \p_{x}^{k-1} \hat{Y}(q)+\dots,\label{KY}
\end{align}
where ``$\dots$" denotes some additional terms that contain $\p_{x}^{i}\hat{L}(q)$ with $i \geq 1$. These terms do not contribute to the expressions of the central invariants of the bihamiltonian structure, so we will omit them. For the same reason, we can replace $\hat{L}(q)$ by $\lambda(x,q)$ in the above formula for $K_Y$.

Substituting \eqref{SymX}, \eqref{SymY} and \eqref{KY} into the above formulae for the two Poisson brackets we obtain 
\begin{align}\label{Main1}
&\{\lm(y,\xi), \lm(z,\zeta)\}_a\nn\\
=& \int dx \oint \frac{dp}{2 \pi i} \oint \frac{dq}{2 \pi i} \sum_{r,s \geq 0} \epsilon^{r} 
\left[
f(p) R^{1}_{r,s;a}(p,q) \partial_x^{r-s+1} g(q) \right.\nn\\
&\quad +  f(p) R^{2}_{r,s;a}(p,q)  \partial_x^{r-s+1} b(x) 
+ a(x) R^{3}_{r,s;a}(p,q) \partial_x^{r-s+1} g(q) \nn\\
&\quad \left.+  a(x) R^{4}_{r,s;a}(p,q) \partial_x^{r-s+1} b(x)\right] ,\quad a=1, 2.
\end{align}
Here integration by parts w.r.t. the variables $x, p, q$ is performed. A straightforward computation gives the following expressions of the first few coefficients of $R^i_{r,s;a}(p,q)$: 
\begin{align*}
R^{1}_{0,0;1} =& \frac{\lambda'(p) - \lambda'(q)}{p-q},\quad R^{2}_{0,0;1} = R^{3}_{0,0;1} = R^{4}_{0,0;1} = 0, \\
R^{1}_{1,0;1} =& \frac{\lambda''(p) + \lambda''(q)}{2(p-q)} - \frac{\lambda'(p) - \lambda'(q)}{(p-q)^2}, \\
R^{2}_{1,0;1} =& -R^{3}_{1,0;1} = -\frac{\lambda '(p)-\lambda '(q)}{p-q}, \quad R^{4}_{1,0;1} = 0, \\
R^{1}_{2,0;1} =& \frac{\lambda'''(p) - \lambda'''(q)}{6(p-q)} -  \frac{\lambda''(p) + \lambda''(q)}{2(p-q)^2} + \frac{\lambda'(p) - \lambda'(q)}{(p-q)^3}, \\
R^{2}_{2,0;1} =& R^{3}_{2,0;1} =- \frac{\lambda ''(p) - \lambda ''(q)}{2(p-q)},\; R^{4}_{2,0;1} = -\frac{\lambda'(p) - \lambda'(q)}{p-q},
\end{align*}
and
\begin{align*}
R^{1}_{0,0;2} =& \frac{\lambda (q) \lambda '(p) - \lambda(p) \lambda '(q)}{p-q} + \frac{\lambda '(p) \lambda '(q)}{n+1}, \\
R^{2}_{0,0;2} =& R^{3}_{0,0;2} = R^{4}_{0,0;2} = 0, \\
R^{1}_{1,0;2} =& \frac{\lambda (p) \lambda '(q)-\lambda (q) \lambda'(p)}{(p-q)^2}+\frac{\lambda (q) \lambda ''(p)+\lambda (p) \lambda ''(q)-2 \lambda '(p) \lambda '(q)}{2 (p-q)} \\
 &\; + \frac{\lambda ''(p) \lambda '(q)-\lambda '(p) \lambda ''(q)}{2 (n+1)} , \\
R^{2}_{1,0;2} =& -R^{3}_{1,0;2} = \frac{\lambda (p) \lambda '(q)-\lambda(q) \lambda '(p)}{p-q} - \frac{\lambda '(p) \lambda '(q)}{n+1}, \quad R^{4}_{1,0;2} = 0, \\
R^{1}_{2,0;2} =& \frac{\lambda'''(p)\lambda(q) - 3\lambda''(p)\lambda'(q) + 3\lambda'(p)\lambda''(q) - \lambda(p)\lambda'''(q)}{6(p-q)}  \\
 &\; - \frac{\lambda''(p) \lambda(q) - 2\lambda'(p) \lambda'(q)  + \lambda(p)\lambda''(q)}{2(p-q)^2} + \frac{\lambda'(p) \lambda(q) - \lambda'(q)\lambda(p)}{(p-q)^3} \\
 &\; + \frac{2\lambda'''(p)\lambda'(q) - 3\lambda''(p)\lambda''(q) + 2\lambda'(p)\lambda'''(q)}{12(n+1)}, \\
R^{2}_{2,0;2} =&\frac{\lambda (p) \lambda ''(q)-\lambda (q) \lambda ''(p)}{2 (p-q)}-\frac{\lambda ''(p) \lambda '(q)+\lambda '(p) \lambda ''(q)}{2 (n+1)} , \\
R^{3}_{2,0;2} =&\frac{\lambda (p) \lambda ''(q)-\lambda (q) \lambda ''(p)}{2 (p-q)}+\frac{\lambda ''(p) \lambda '(q)-\lambda '(p) \lambda ''(q)}{2 (n+1)} , \\
R^{4}_{2,0;2} =& \frac{\lambda (p) \lambda '(q)-\lambda(q) \lambda '(p)}{p-q} - \frac{\lambda '(p) \lambda '(q)}{n+1}.
\end{align*}
Here and in what follows $\lm'(p)$, $\lm''(p)$, $\lm'''(p)$ denote the functions $\p_p \lm(x,p)$, $\p_p^2\lm(x,p)$, $\p_p^3\lm(x,p)$  respectively. It is easy to check that all the r.h.s. of the above formulae have no singularity at $p=q$. 

\begin{lemma}
The Poisson brackets \eqref{Main1} can be represented in the form 
\begin{align*}
\{\lambda (y,\xi) ,\lambda(z,\zeta)\}_a = \sum_{k \geq 0} \sum_{l=0}^{k+1} \epsilon ^k A_{k,l;a}(\xi,\zeta)\,\delta^{(k-l+1)}(y-z),\quad a=1,2.
\end{align*}
Here the coefficients $A_{k,l;a}(\xi,\zeta)$ that are relevant to our computation of the central invariants have the expressions
\begin{align*}
A_{0,0;1} &= R^{1}_{0,0;1}(\xi , \zeta), \\
A_{1,0;1} &= R^{1}_{1,0;1}(\xi , \zeta) - \frac{(\xi - \zeta) (\xi + \zeta - 2w(x)) u(x)}{(\xi - w(x))^3(\zeta - w(x))^3 },\\
A_{2,0;1} &= R^{1}_{2,0;1}(\xi , \zeta) - \frac{(\zeta +\xi -2 w(x)) \left(\zeta ^2+\xi ^2-\xi \zeta +w(x)^2-\zeta  w(x)-\xi w(x)\right) u(x)}{(\zeta -w(x))^4 (\xi -w(x))^4},\\
A_{0,0;2} &= R^{1}_{0,0;2}(\xi , \zeta),\\
A_{1,0;2} &= R^{1}_{1,0;2}(\xi , \zeta) - \frac{(\xi + \zeta - 2w(x)) u(x)}{(\zeta - w(x))^2 (\xi - w(x))^2} \left(\frac{\lambda (\xi)}{\zeta -w(x)}-\frac{\lambda (\zeta )}{\xi -w(x)}\right) \\
 &\quad - \frac{u(x)}{(\zeta-w(x)) (\xi-w(x))} \left(\frac{\lambda '(\xi )}{\zeta -w(x)}-\frac{\lambda'(\zeta )}{\xi -w(x)}\right) \\
 &\quad + \frac{u(x)}{n+1}\left(\frac{\lambda '(\xi)}{(\zeta -w(x))^3}-\frac{\lambda '(\zeta )}{(\xi-w(x))^3}\right),\\
A_{2,0;2} &= R^{1}_{2,0;2}(\xi , \zeta) - \frac{n+2}{n+1} \frac{u(x)^2}{(\zeta -w(x))^3 (\xi -w(x))^3}  \\
&\quad - \frac{(\xi -\zeta ) (\zeta +\xi -2w(x)) u(x)}{(\zeta -w(x))^3 (\xi -w(x))^3}\left(\frac{\lambda (\xi )}{\zeta -w(x)}-\frac{\lambda (\zeta )}{\xi-w(x)}\right) \\
&\quad - \frac{u(x)}{2 (\zeta -w(x)) (\xi -w(x))}\left(\frac{\lambda ''(\zeta )}{\xi -w(x)}+\frac{\lambda ''(\xi )}{\zeta-w(x)}\right) \\
&\quad - \frac{u(x)}{(\zeta -w(x)) (\xi -w(x))}\left(\frac{\lambda '(\zeta )}{(\xi-w(x))^2}+\frac{\lambda '(\xi )}{(\zeta -w(x))^2}\right) \\
&\quad - \frac{2 u(x) \lambda _+'(w(x))}{(\zeta -w(x))^2 (\xi -w(x))^2}-\frac{u(x)(\zeta +\xi -2 w(x))\lambda _+(w(x)) }{(\zeta -w(x))^3 (\xi -w(x))^3} \\
&\quad + \frac{u(x)}{n+1}\left(\frac{\lambda ''(\zeta )}{2 (\xi-w(x))^3}+\frac{\lambda ''(\xi )}{2 (\zeta -w(x))^3}+\frac{\lambda '(\zeta )}{(\xi-w(x))^4}+\frac{\lambda '(\xi )}{(\zeta -w(x))^4}\right).
\end{align*}
In the above formula $\lm_+(p) = \lm(x, p)-\frac{u(x)}{p-w(x)}$ denotes the positive part of the symbol $\lm(p)$, and $\lm_+'(p)=\p_p \lm_+(p)$.
\end{lemma}

\begin{prf}
In order to simplify the computation,  let us introduce the following function:
\begin{align*}
\bar{f}(p) 
=& f(p) + \sum_{i>n} \frac{\xi^{i-1}}{p^i} \delta(x-y) \\
=& \sum_{i>0} \frac{\xi^{i-1}}{p^i} \delta(x-y) + a_0(x) + p\,a_{-1}(x) \\
=& \frac{1}{p-\xi}\delta(x-y) +  a_0(x) + p\,a_{-1}(x),
\end{align*}
where we assume that $\xi$ is a complex number such that $|p|>|\xi|$. In the same way we introduce the function 
\begin{align*}
\bar{g}(q) = \frac{1}{q-\zeta}\delta(x-z) +  b_0(x) + p\,b_{-1}(x).
\end{align*}
It is easy to see that if we replace $f(p)$ and $g(q)$ in the r.h.s. of \eqref{Main1} by $\bar{f}(p)$ and $\bar{g}(q)$ respectively, the r.h.s. of the expressions of the Poisson brackets are unchanged. After this replacement the path integral w.r.t.  $p$ and $q$ can be computed by taken the residues at $p=\xi$, $p=w$, $q=\zeta$ and $q=w$. A direct computation leads to the results of the lemma. The lemma is proved.
\end{prf}

Now we are ready to compute the functions $f^{i}$, $P^{ij}_a$ and $Q^{ii}_a$ that 
appear in the definition \eqref{CenInv} of the central invariants. Suppose $r_i\;(i=1,2,\ldots, n+2\,)$ are the roots of $\lambda'(r)=0$ and pairwise distinct, then it is easy to see that $\lambda(r_i) \;(i=1,2,\ldots, n+2)$ provide a system of canonical coordinates of the bihamiltonian structure \eqref{PoBkt1}, \eqref{PoBkt2}. It follows from the above lemma that 
\begin{align*}
f^i =& \lambda''(r_i), \\
P^{ij}_1 =& \frac{\lambda''(r_i) + \lambda''(r_j)}{2(r_i - r_j)} - \frac{(r_i-r_j)(r_i+r_j-2w) u}{(r_i-w)^3(r_j-w)^3}, \\
P^{ij}_2 =& \frac{\lambda''(r_i) \lambda(r_j) +  \lambda(r_i)\lambda''(r_j)}{2(r_i - r_j)} \\
& \qquad - \frac{(r_i+r_j-2w)(\lambda(r_i)(r_i-w)-\lambda(r_j)(r_j-w)) u}{(r_i-w)^3(r_k-w)^3}, \\
Q^{ii}_1 =& \frac{1}{12} \lambda''''(r_i) - \frac{2u}{(r_i-w)^5}, \\
Q^{ii}_2 =& \frac{1}{12} \lambda''''(r_i)\lambda(r_i) + \frac{n}{4(n+1)}\lambda''(r_i)^2 - \frac{n+2}{n+1}\frac{u^2}{(r_i-w)^6} \\
& \quad - \frac{n}{n+1} \frac{u\lambda''(r_i)}{(r_i-w)^3}-\frac{2 u \lambda_+'(w)}{(r_i-w)^4} - \frac{2u \lambda _+(w)}{(r_i-w)^5},
\end{align*}
for $i,j=1,2,\ldots,n+2$. 
From the above formulae we obtain 
\begin{align*}
 &P_{2}^{ki}-\lm(r_i)P_{1}^{ki} \\
=&  \left( \lm(r_k) - \lm(r_i) \right) \left[ \frac{\lm''(r_i)}{2(r_k - r_i)} - \frac{u(r_k+r_i-2w)}{(r_k-w)^3 (r_i-w)^2} \right] \\
=&  \frac{\lm(r_k) - \lm(r_i)}{r_k - r_i}\left[ \frac{\lm_+''(r_i)}{2} + \frac{u}{(r_i-w)(r_k-w)^2}\right].
\end{align*}
So we have 
\begin{align*}
& \sum_{k\neq i} \frac{(P_2^{ki}-\lm(r_i)P_1^{ki}))^2}{f^k(\lm(r_k) - \lm(r_i))} \\
=&  \frac{\lm_+''(r_i)^2}{4} \sum_{k \neq i} \frac{\lm(r_k)-\lm(r_i)}{\lm''(r_k)(r_k -r_i)^2}  + \frac{u \lm_+''(r_i) }{r_i-w} \sum_{k \neq i} \frac{\lm(r_k)-\lm(r_i)}{\lm''(r_k)(r_k -r_i)^2(r_k-w)^2} \\
& \qquad + \frac{u^2}{(r_i-w)^2} \sum_{k \neq i} \frac{\lm(r_k)-\lm(r_i)}{\lm''(r_k)(r_k -r_i)^2(r_k-w)^4} \\
=& \frac{1-n}{2(n+1)} \frac{\lm''(r_i)^2}{4} + \frac{u \lm''(r_i) }{(r_i-w)^3} + \frac{n+2}{n+1}\frac{u^2}{(r_i-w)^6} \\
& \qquad  -\frac{2u \lm(r_i)}{(r_i-w)^5} + \frac{2 u \lm_+(w)}{(r_i-w)^5} + \frac{2 u\lm_+'(w)}{(r_i-w)^4}. 
\end{align*}
To obtain the last equality we applied the residue theorem to the following meromorphic function:
\[h_1(z)= \frac{\lambda(z)-\lambda(r_i)}{\lambda'(z) (z-r_i)^2},\quad 
h_2(z) = \frac{h_1(z)}{(z-w)^2},\quad h_3(z) = \frac{h_1(z)}{(z-w)^4}.\]
From the above expressions for the functions $Q_1^{ii}$ and $Q_2^{ii}$ we also have  
\begin{align*}
& Q_2^{ii}-\lm(r_i)Q_1^{ii} = \frac{n}{n+1} \frac{\lm''(r_i)^2}{4} - \frac{u \lm''(r_i)}{(r_i-w)^3} - \frac{n+2}{n+1}\frac{u^2}{(r_i-w)^6} \\
& \qquad \qquad + \frac{2u\lm(r_i)}{(r_i-w)^5} - \frac{2u \lm_+(w)}{(r_i-w)^5} - \frac{2u\lm_+'(w)}{(r_i-w)^4}.
\end{align*}
So by using the definition \eqref{CenInv} we obtain
\begin{align*}
c_i = \frac{1}{3 \lm''(r_i)^2} \left[ \frac{n}{n+1}\frac{\lm''(r_i)^2}{4} + \frac{1-n}{2(n+1)} \frac{\lm''(r_i)^2}{4} \right] =\frac{1}{24} .
\end{align*}
Thus we proved  the Main Theorem. 

\section{Conclusion}\label{sec-5}
We have shown that the central invariants of the bihamiltonian structure \eqref{PoBkt1}, \eqref{PoBkt2} of the constrained
KP hierarchy \eqref{jw-1} are equal to $\frac1{24}$, which is an important feature of the topological deformation of the principal hierarchy of any semisimple Frobenius manifold. There is indeed an
$(n+2)$-dimensional semisimple Frobenius manifold $M$ underlined the constrained KP hierarchy. It has the superpotential 
\[\lm(p)=p^{n+1}+v_n p^{n+1}+\dots+v_2 p+v_1+\frac{u}{p-w},\] 
and the Frobenius manifold structure is defined as follows (see \cite{Du} for details). The flat metric of the Frobenius manifold is defined by the residue pairing
\[ \langle \frac{\p}{\p v_i}, \frac{\p}{\p v_j}\rangle
=-(\res_{p=\infty}+\res_{p=w}) \frac{\p_{v_i} \lm(p)\p_{v_j}\lm(p) dp}{\p_p \lm(p)},
\]
and the multiplication is defined by 
\[ c(\frac{\p}{\p v_i}, \frac{\p}{\p v_j}, \frac{\p}{\p v_k})
=-(\res_{p=\infty}+\res_{p=w}) \frac{\p_{v_i} \lm(p)\p_{v_j}\lm(p) \p_{v_k} \lm(p)dp}{\p_p \lm(p)}.
\]
Here $i, j=1,\dots, n+2$, and we denote $v_{n+1}=w, v_{n+2}=u$. The flat coordinates 
of the Frobenius manifold can be chosen as 
\begin{align*}
&\tilde{v}_i=-\frac{n+1}{n+1-i} \res_{p=\infty} \lm(p)^{1-\frac{i}{n+1}},\quad i=1,\dots,n,\\
&\tilde{v}_{n+1}=v_{n+1},\quad \tilde{v}_{n+2}=v_{n+2}.
\end{align*}
Then potential $F(\tilde{v})$ of the Frobenius manifold can be represented as 
\[ F=P(\tilde{v}_1,\dots,\tilde{v}_{n+2})+\frac12 v_{n+2}^2 \left(\log \tilde{v}_{n+2}-\frac32\right),\]
where $P(\tilde{v})$ is a quasi-homogeneous polynomial which does not contain quadratic 
monomials, and the quadratic term $-\frac34 \tilde{v}_{n+2}^2$ is added to the potential for the convenience of fixing a calibration 
\[ \{\theta_{i,m}(\tilde{v})\,|\, m\ge 0, i=1,\dots, n+2\}\]
of the Frobenius manifold as it is done 
in \cite{LXZ}. 

The flat metric $\langle\,,\rangle$ defines a Poisson bracket of hydrodynamic type on the loop space of the Frobenius manifold $M$, and it is easy to see that this Poisson bracket coincides with the one obtained from the first Poisson bracket \eqref{PoBkt1} of the 
constrained KP hierarchy \eqref{jw-1} under the dispersionless limit $\epsilon\to 0$ \cite{Du}.
There is a second flat metric of the Frobenius manifold which is define outside of the discriminant of $M$ and is given by the following residue pairing:
\[ ( \frac{\p}{\p v_i}, \frac{\p}{\p v_j})
=-(\res_{p=\infty}+\res_{p=w}) \frac{\p_{v_i} \lm(p)\p_{v_j}\lm(p)  dp}{\lm(p) \p_p \lm(p)},\quad i, j=1,\dots, n+2.
\]
It yields another Poisson bracket of hydrodynamic type on the loop space of the Frobenius manifold $M$, and it coincides with the one obtained from the second Poisson bracket \eqref{PoBkt2} of the 
constrained KP hierarchy \eqref{jw-1} under the dispersionless limit. These two Poisson brackets 
yield a bihamiltonian structure on the loop space of the Frobenius manifold and a bihamiltonian integrable hierarchy of hydrodynamic type -- called the principal hierarchy of the Frobenius manifold. It can be represented as 
\begin{equation}\label{ph-1}
\frac{\p \tilde{v}_i}{\p t^{j,m}}=\eta^{ik} \frac{\p}{\p x}\left(\frac{\p\theta_{j,m+1}}{\p \tilde{v}_k}\right),\quad m\ge 0,\ i, j=1,\dots, n+2,
\end{equation}
where the matrix $(\eta^{ij})$ is given by the inverse of the matrix $(\langle \frac{\p}{\p \tilde{v}_i}, \frac{\p}{\p \tilde{v}_j}\rangle)$,
and the functions $\theta_{j,m}$ which defines the calibration of the Frobenius manifold can be represented as
\begin{align*}
\theta_{j,m}&=-\frac1{(n+1) \prod_{k=0}^{m} (k+\frac{j}{n+1})} \res_{p=\infty}
\lm(p)^{m+\frac{j}{n+1}},\quad m\ge 0,\  j=1,\dots,n.\\
\theta_{n+1,m}&=-\frac1{(m+1)!} \res_{p=\infty}
\lm(p)^{m+1},\quad m\ge 0.\\
\theta_{n+2,m}&=\frac{n+2}{(n+1) m!}\res_{p=w}\left[ \lm(p)^m \left(\log\lm(p)-c_m\right)\right],\quad m\ge 0.
\end{align*}
Here $c_0=0, c_m=\sum_{k=1}^m \frac1{k}$, and $\log\lm(p)$ is redefined as follows so that we can take the residue near $p=w$. We first expand the function 
\[ \log\left(\frac{p-w}{u} \lm(p)\right)=\log\left(1+\sum_{k=1}^{n}\frac{v_k}{v_{n}}p^{k-1} (p-w)+
\frac{1}{u}p^{n+1} (p-w)\right)\]
near $p=w$ to obtain the formal power series
\[A=\sum_{k\ge 1} a_k (p-w)^k.\]
Then we consider the function 
\[\log\left(\frac{1}{(p-w)^{n+1}}\lm(p)\right)\]
and view it as a function of $\tilde{p}=p-w$. We expand it near $\tilde{p}=\infty$ 
to obtain the following formal power series of $(p-w)^{-1}$:
\[B=\sum_{k\ge 1} b_k (p-w)^{-k}.\]
Now the logarithm of $\lm(p)$ is define by 
\[\log\lm(p):=\frac{n+1}{n+2} \left(\log{u}+A+\frac1{n+1} B\right).\] 

Denote by $\frac{\p}{\p t_k}, k\ge 1$ the dispersionless limit of the constrained
KP hierarchy \eqref{jw-1}, then they have the following relation with the principal
hierarchy \eqref{ph-1} of the Frobenius manifold $M$:
\begin{align*}
\frac{\p}{\p t^{j,m}}&=\frac1{n+1}\prod_{k=0}^{m-1} (k+\frac{j}{n+1})^{-1} \frac{\p}{\p t_{m (n+1)+j}},\quad m\ge 0, j=1,\dots,n.\\
\frac{\p}{\p t^{n+1,m}}&=\frac1{(m+1)!}\frac{\p}{\p t_{m+1}},\quad m\ge 0.
\end{align*}
In order to establish the relation of the flows $\frac{\p}{\p t^{n+2,m}}$ with that of the constrained KP hierarchy, we need to add to the constrained KP hierarchy another set of flows as it was done 
for the $1$-constrained KP hierarchy in  \cite{CDZ}. The main step in the construction of the additional flows given in \cite{CDZ} is to define the logarithm of the Lax operator $L$. 
This was achieved by a Legendre transformation that relates the $1$-constrained KP hierarchy with the extended Toda hierarchy. This construction can be generalized to
give a Legendre transformation between the constrained KP hierarchy and a special
case of the bigraded Toda hierarchy given in \cite{Ca}, and to give a definition of the logarithm 
of the Lax operator $L$. After an appropriate definition of the additional set of flows of the 
constrained KP hierarchy, we obtain the extended constrained KP hierarchy. We 
have the flowing conjecture:
\begin{conj}
The extended constrained KP hierarchy coincides with the topological deformation of the principal hierarchy of the associated Frobenius manifold.
\end{conj}
We note that a canonical way is given in \cite{DZ} to obtain the topological deformation of the principal hierarchy associated to a semisimple Frobenius manifold. We will study the 
validity of the above conjecture in a separate publication. 
Finally, we remark that in \cite{CDZ2} it is proved that the central invariants of the bihamiltonian structure of the bigraded Toda hierarchy are also equal to $\frac1{24}$. It would be interesting to establish a direct relation between the bihamiltonian structures of the constrained KP hierarchy and that of a certain special bigraded Toda hierarchy.

\vskip 0.5truecm 
\paragraph{Acknowledgments}
The authors thank Chao-Zhong Wu for helpful discussions.
This work  is partially supported by NSFC No.\,11171176, No.\,11222108, No.\,11371214, and  No.\,11471182.

Emails:
\begin{itemize}
\item[] Si-Qi Liu: liusq@tsinghua.edu.cn
\item[] Youjin Zhang: youjin@tsinghua.edu.cn
\item[] Xu Zhou: x-zhou09@mails.tsinghua.edu.cn
\end{itemize}

\end{document}